\documentstyle[mnras_cite,psfig]{mn}

\def\lsim{ \lower .75ex \hbox{$\sim$} \llap{\raise .27ex \hbox{$<$}} }
\def\gsim{~\rlap{$>$}{\lower 1.0ex\hbox{$\sim$}}}
\def\d{{\rm d}}
\begin{document}

\title{The evolution of the galaxy distribution} 
\author[A.~J.~Benson, C.~S.~Frenk, C.~M.~Baugh, S.~Cole,  \&
C.~G.~Lacey]{A.~J.~Benson$^{1,3,4}$, C.~S.~Frenk$^1$, C.~M.~Baugh$^1$, S.~Cole$^1$  \& C.~G.~Lacey$^2$ \\
1. Physics Department, University of Durham, Durham, DH1 3LE, England \\
2. SISSA, via Beirut 2-4, 34014 Trieste, Italy \\
3. Present address: California Institute of Technology, MC 105-24, Pasadena, CA 91125-2400, USA \\
4. E-mail: abenson@astro.caltech.edu}

\maketitle

\begin{abstract}

We follow the evolution of the galaxy population in a $\Lambda$CDM
cosmology by means of high-resolution N-body simulations in which the
formation of galaxies and their observable properties are calculated
using a semi-analytic model. We display images of the spatial
distribution of galaxies in the simulations that illustrate its
evolution and provide a qualitative understanding of the processes
responsible for various biases that develop. We consider three
specific statistical measures of clustering at $z=1$ and $z=0$: the
correlation length (in real- and redshift-space) of galaxies of
different luminosity, the morphology-density relation and the genus
curve of the topology of galaxy isodensity surfaces. For galaxies with
luminosity below $L_*$, the $z=0$ correlation length depends very
little on the luminosity of the sample, but for brighter galaxies it
increases very rapidly, reaching values in excess of
$10h^{-1}$Mpc. The ``accelerated'' dynamical evolution experienced by
galaxies in rich clusters, which is partly responsible for this
effect, also results in a strong morphology-density
relation. Remarkably, this relation is already well-established at
$z=1$. The genus curves of the galaxies are significantly different
from the genus curves of the dark matter but this is not due to
genuine topological differences but rather to the sparse sampling of
the density field provided by galaxies.  The predictions of our model
at $z=0$ will be tested by forthcoming data from the 2dF and Sloan
galaxy surveys, and those at $z=1$ by the DEEP and VIRMOS surveys.
\end{abstract}

\section{Introduction}

Studies of the large-scale distribution of galaxies have traditionally
focussed on problems such as testing hypotheses for the identity of
the dark matter, the nature of the initial density perturbations and
the mechanism of structure growth. Properties of the observed
large-scale structure are also often used to estimate the values of
fundamental cosmological parameters. Although none of these issues can
be regarded as settled, there is now a growing consensus that cold
dark matter (CDM) is the most likely candidate for the dark matter,
that cosmic structure grew by the gravitational amplification of
random-phase initial density fluctuations of inflationary origin, and
that the fundamental cosmological parameters have the following
values: density parameter, $\Omega_0 \simeq 0.3$, cosmological
constant term, $\Lambda_0 \simeq 0.7$ and Hubble constant (in units of
100 km s$^{-1}$ Mpc$^{-1}$), $h\simeq 0.7$.

Cosmological constraints reflect only one aspect of the information
encoded in the pattern of galaxy clustering. Another, equally
interesting aspect, concerns the processes responsible for the
formation and evolution of galaxies. To extract this kind of
information requires very extensive datasets and these are only now
becoming available in the form of a new generation of galaxy surveys,
like 2dF \cite{peacock2001}, Sloan \cite{blanton00} and 2MASS
\cite{jarret00}. The expectation is that these new datasets will
provide, in addition to further cosmological constraints, some
understanding of how the physics of galaxy formation manifests itself
in the clustering of galaxies as a function of internal properties
such as morphology, luminosity, colour, or star formation rate. Not
only is this important for testing models of galaxy formation, but it
is also required for extracting accurate cosmological information from
the new surveys. Although it seems plausible that on very large scales
the galaxy distribution traces the underlying mass in a simple way
\cite{coles93,colehatton98}, complex biases are predicted to be
present on small and intermediate scales \cite{kauff99a,meetal}.

To extract useful information from observational data of the quality
and size of the new surveys, it is necessary to have detailed
theoretical predictions. In this area too there have been significant
advances in recent years, largely through the development of
increasingly realistic {\it ab initio} calculations of galaxy formation
and evolution. Two strategies have been developed for this purpose. In
the first one, cosmological N-body/gasdynamics simulations are used to
follow the coupled evolution of dark matter and gas, in particular,
the cooling of gas in galactic dark matter halos
(e.g. \pcite{katz92,evrard94,frenk96,weinberg97,blanton99,pearce99}). A
phenomenological model is employed to decide when and where stars and
galaxies form from this cooled gas and to include the associated
feedback effects. In the second strategy, only the evolution of the
dark matter component is simulated directly, or the assembly history
of halos is obtained with a 
Monte-Carlo method, and the behaviour of the gas is calculated by
solving a simple, analytical, spherically symmetric cooling-flow
model. As in the direct simulation approach, star formation and
feedback are included in a phenomenological way.

The two strategies offer different advantages. Direct
simulations solve the evolution equations for
gravitationally coupled dark matter and dissipative gas without
imposing any restrictions on geometry. However, limited resolution
restricts the range of length and mass scales that can be studied, 
and the expense of large
simulations makes it impractical to carry out extensive parameter
space explorations. Because of its simplified treatment of gas
dynamics, semi-analytic modelling can follow an essentially unlimited
range of length and mass scales, and is sufficiently flexible that the effects
of varying assumptions and parameter values can be readily
explored. Additional processes that cannot currently be easily investigated
at the resolution available in direct simulations, such as those
determining galaxy
morphology, or the effects of dust obscuration, can be readily
incorporated into the semi-analytic models by straightforward
extensions to the phenomenological model of star formation and
feedback. The numerical resolution and physical content of a typical
N-body/gasdynamic simulation can be mimicked in a semi-analytic model
and \scite{benson00d} have shown that, at least in the case where only
the simplest gas physics are modelled, the two techniques give
reassuringly similar statistical results.

In this paper we combine large N-body simulations with the
semi-analytic model of \scite{coleetal} to investigate certain
properties of the galaxy distribution that are relevant to the new
generation of galaxy redshift surveys. We begin by displaying images
that illustrate the evolution of the galaxy population in a
representative volume of a simulated CDM universe. These images
furnish some qualitative understanding of the mechanisms responsible
for establishing the spatial distribution of galaxies of different
kinds. We then focus specifically on the dependence of the two-point
correlation function on galaxy luminosity, the morphology-density
relation, and the topology of the galaxy distribution as measured by
the genus, and on the evolution of these properties with redshift. The
dependence of clustering on luminosity and colour have previously been
considered, using similar techniques, by \scite{kns,kauff99a} and
\scite{benson00c}. The first and last of these studies found a weak
increase in the correlation function with luminosity on large scales,
but the second failed to detect any effect.  These papers used
simulations of relatively small volumes and so were unable to
investigate the clustering of the brightest galaxies, for which
luminosity-dependent effects are expected to be strongest. The new
generation of redshift surveys will include large samples of very
bright galaxies and may well be able to measure this kind of
effect. In this paper we extend earlier work and investigate
clustering at the bright end of the galaxy luminosity
function. Closely related to the dependence of clustering strength on
luminosity is the morphology-density relation which we also quantify
in our simulations, both at the present day and at $z=1$. Finally, we
provide the first determination of the genus curves predicted for {\it
galaxies} in a CDM model; previous simulations had only been able to
address the genus curves of the dark matter distribution. Although our
model predictions are directed at the new surveys, we carry out
limited comparisons with available observational data.

The remainder of this paper is organised as follows. In \S2, we
describe our simulation and modelling techniques. In \S3 we present
colour images of the evolution of a slice of our simulated volume
(these images are publically available at {\tt
\verb+http://www.astro.caltech.edu/~abenson/Mocks/mocks.html+}). In
\S4 we present quantitative estimates of clustering, namely the
correlation length-luminosity and morphology-density relations and the
genus statistic, and compare our results to observations. Finally, we
present our conclusions in \S5.

\section{Method}

The need for realistic modelling of galaxy formation as a prerequisite
for deriving reliable clustering predictions has been emphasised by
\scite{meetal} who showed, for example, that the form of the two-point
correlation function on small scales is strongly influenced by the
physical processes governing galaxy formation. Such processes are
readily taken into account when the techniques of semi-analytic
modelling are grafted into N-body simulations of the dark matter
\cite{kns,kauffdummy,diaferio99,ajbdummy,somerville01a}. Monte-Carlo
implementations of the semi-analytic model can also be used directly
for clustering studies, without N-body simulations, but they only work
well on scales larger than the Lagrangian radii of the dark matter
halos which host galaxies, in practice on scales in excess of a few
Mpc (although see \pcite{seljak00}), for which the bias can be
calculated using the analytic formula of \scite{mowhite96}; see for
example \scite{baugh99}. Modelling the visible properties of galaxies
explicitly allows simulated samples to be selected according to
criteria closely patterned on observational selection procedures
(e.g. by magnitude, colour, morphology, etc), thus allowing rigorous
comparisons with observations to be made.
 
In this paper, we use the techniques introduced by \scite{kns} and
extended by \scite{meetal} to graft our semi-analytic model of galaxy
formation onto N-body simulations.  Full details of our semi-analytic
model and the extensions required to study galaxy clustering are given
in \scite{coleetal} and \scite{ajbdummy} respectively. Briefly, dark
matter halos are identified in the simulation at the redshift of
interest using the friends-of-friends algorithm with the standard
linking length of 0.2 \cite{davis85}. After cleaning the halo
catalogue in the manner described below, the mass of each halo is
input into the semi-analytic model. By means of simple,
physically-motivated prescriptions, described in detail in
\scite{coleetal}, the model calculates the amount of gas that cools in
a virialized halo of that mass to make a galaxy, as well as the star
formation rate, the reheating of left-over gas by stellar winds and
supernovae, and the chemical evolution of the gas and stars. Galaxies
are allowed to merge within common dark matter halos, producing
elliptical galaxies and bulges from stellar disks. The
spectrophotometric evolution of the galaxies is calculated using a
standard stellar population synthesis model
\cite{bc93,bc01}. Extinction by dust is included using the models of
\scite{ferrara99}. The most massive galaxy in each halo is identified
with the central galaxy and placed at the centre of mass (and given
the peculiar velocity of the centre of mass). Other galaxies
(satellites) are assigned the position and peculiar velocity of a
randomly chosen dark matter particle within the halo. In this way,
satellite galaxies always trace the dark matter within a given halo.
Modelling of this kind has been successfully applied to study a large
variety of properties of the galaxy population (e.g
\pcite{coleetal94,kgw94,baugh96dummy,kauff96,baugh98,somerville99,coleetal,granato00}.)

We consider only dark matter halos in the simulation containing ten or
more particles. Since we are interested in galaxies of all
luminosities, including the faint ones that occupy halos with masses
close to the ten-particle limit, it is important to check that small
halos are actually bound objects.  For this, we calculate the total
energy of each halo by summing the kinetic energy (measured relative
to the centre of mass of the halo) and the gravitational energy (due
to the interaction between all of the particles in the halo). If a
halo is found to have positive energy (and so to be unbound), we
remove the least bound particle and recompute the total energy. This
process is repeated until either the energy becomes negative (in which
case we now have a bound halo with a lower mass than the original) or
there are fewer than ten particles left in the halo (in which case we
discard it). In this way, we construct a new halo catalogue containing
only bound objects.

Typically, approximately 10\% of the halos from the original catalogue
fail the binding energy test and are excluded. A slightly smaller
fraction have particles removed but remain in the catalogue. Most of
the excluded halos come from the low mass end of the distribution,
with the excluded fraction dropping rapidly as the halo mass
increases. We find many examples of halos contaminated by interlopers
for which removal of a small number of the least bound particles
results in a bound object. (Note that since the binding energy test
only affects halos near the resolution limit of the N-body simulation,
it does not alter any of the results of \scite{meetal} who considered
only bright galaxies that form in halos well above the resolution
limit.)

We adopt similar values for the parameters of the semi-analytic model
as did \scite{coleetal}, except that the parameters describing the
normalisation and shape of the power spectrum ($\sigma_8$ and
$\Gamma$, see below) are set to the actual values used in the
simulation. The parameters of the semi-analytic model 
are slightly different from
those used in \scite{meetal}, but the differences in the predictions are
negligible. Furthermore, since the dark matter halo mass function in
the simulations differs somewhat from the Press-Schechter form assumed
by \scite{coleetal}, we find that the model works better if the value
of $\Upsilon$ (the ratio of total to visible stellar mass, which
depends on the fraction of brown dwarfs) is set
equal to 1 rather than 1.4 as in \scite{coleetal}. The recycled
fraction in the calculation of the chemical enrichment is modified
accordingly \cite{coleetal}. \scite{meetal} showed that in a
$\Lambda$CDM cosmology, this model produces a real-space two-point
galaxy correlation function which is remarkably similar to that
measured in the APM survey by \scite{cmbapm}, in contrast to the
$\tau$CDM cosmology which fails to match the observed two-point
correlation function on all scales. Furthermore, \scite{meetal} showed
that clustering predictions are robust to changes in the semi-analytic
parameters provided that the model matches the bright end of the local
galaxy luminosity function. The evolution of the galaxy correlation
function with redshift is in good agreement with SPH simulations of
galaxy formation \cite{pearce99,benson00d}. In this paper, we consider
only the $\Lambda$CDM model. We have checked that the parameters we
have adopted do produce a correlation function of $L_*$ galaxies
identical to that of \scite{meetal}.

\begin{figure*}
\caption{A slice through the N-body simulation volume at six
redshifts: $z=0.0$, 0.5, 1.0, 2.0, 3.0 and 5.0, as indicated below
each panel. Comoving coordinates are used. The region displayed has
comoving dimensions of $141 \times 141 \times 8 h^{-3}$Mpc$^3$. The
dark matter is represented as a greyscale, with the densest regions
darkest. The positions of the model galaxies are indicated by coloured
circles whose size corresponds to the rest-frame B-band absolute
magnitude of the galaxy, while their colour indicates the rest frame
B-V colour (see the key at the top of the figure). The red and green
boxes indicate regions that are shown in greater detail in
Fig.~\protect\ref{fig:fourzoom}. A high resolution copy of this figure
can be found at {\tt
http://www.astro.caltech.edu/$\sim$abenson/Mocks/mocks.html}}
\label{fig:sixslice}
\end{figure*}

\begin{figure*}
\caption{Slices through selected regions of the N-body simulation
volume at three redshifts: $z=0.0$, 1.0 and 3.0, as indicated below
each panel. Comoving coordinates are used. Each region has comoving
dimensions of $20 \times 20 \times 8 h^{-3}$Mpc$^3$. The dark matter
is represented as a greyscale, with the densest regions darkest. The
positions of the model galaxies are indicated by coloured circles
whose size corresponds to the rest-frame B-band absolute magnitude of
the galaxy while their colour indicates the rest-frame B-V colour of
the galaxy (see the key at the top of the figure). The left-hand
panels show the region around a $z=0$ supercluster (indicated by the
red box in Fig.~\protect\ref{fig:sixslice}), while the right-hand
panels show a more typical region (as indicated by the green box in
Fig.~\protect\ref{fig:sixslice}). A high resolution copy of this
figure can be found at {\tt
http://www.astro.caltech.edu/$\sim$abenson/Mocks/mocks.html}}
\label{fig:fourzoom}
\end{figure*}
 
We use two different N-body simulations. The first one is the ``GIF''
$\Lambda$CDM simulation, a full description of which may be found in
\scite{arj98} and \scite{kauff99a}. This is a 17 million dark matter
particle simulation in a cubic volume of side $141.3 h^{-1}$Mpc, with
cosmological parameters $\Omega _0=0.3$, $\Lambda_0 = 0.7$, $h=0.7$,
$\Gamma = 0.21$ and $\sigma _8 = 0.9$ (where $\Gamma$ is the power
spectrum shape parameter and $\sigma_8$ is the linearly extrapolated
rms mass fluctuation in a sphere of radius $8 h^{-1}$ Mpc). The mass
of the smallest resolved halo in this simulation is $1.4\times
10^{11}h^{-1}M_\odot$.   We also
analyse the ``$512^3$'' simulation described by \scite{arj00} and
\scite{benson00c}, which has identical cosmological parameters to the
GIF simulation (although a slightly different transfer function for
the input power spectrum.)  The particle mass in this simulation is
larger, by a factor of roughly 5, than in the GIF simulation, so that
only dark matter halos more massive than $7\times
10^{11}h^{-1}M_\odot$ are resolved, but the volume is approximately 40
times larger than in the GIF simulation. The $512^3$ simulation is
ideal for studying the brightest galaxies which are only found in very
massive halos and have low abundance. \scite{benson00c} noted that the
dark matter correlation function in the $512^3$ simulation differed
slightly from that in the GIF simulation due to the large-scale power
which is included in the larger volume but is missing in the smaller
one. Throughout this paper, we apply a small correction to all the
correlation functions determined from the GIF simulation by adding the
quantity $\Delta \xi(r)=\xi_{512^3}(r)-\xi_{\rm GIF}(r)$ (where
$\xi_{512^3}$ and $\xi_{\rm GIF}$ are the correlation functions of the
{\it dark matter}, in real or redshift space as appropriate, in the
$512^3$ and GIF simulations respectively). The required correction is
not necessarily the same for galaxies and dark matter of
course. However, the correction is at most 20\% in $\xi(r)$ over the
range of scales considered in this work and, furthermore, ignoring it does not
alter any of our conclusions.

\begin{figure}
\psfig{file=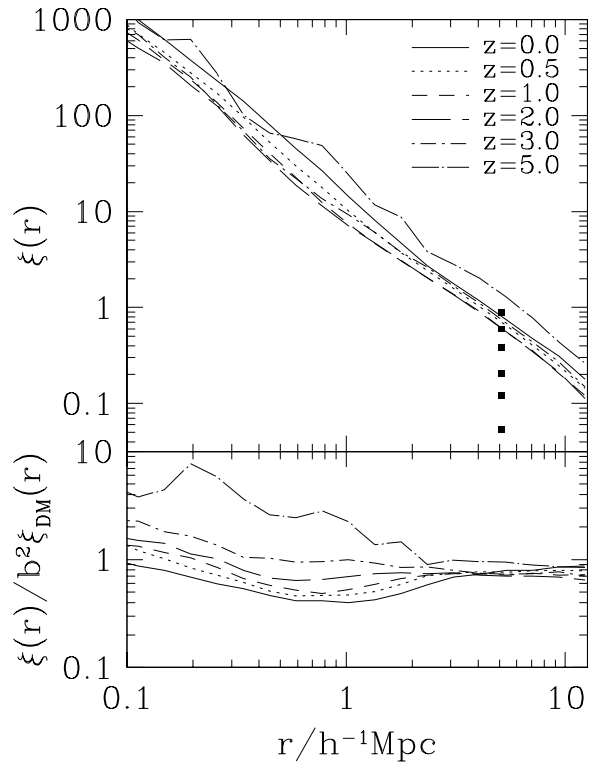,width=80mm,bbllx=0mm,bblly=130mm,bburx=105mm,bbury=265mm,clip=}
\caption{\emph{Top panel:} the real-space correlation functions of
galaxies brighter than rest-frame $M_{\rm B}-5\log h=-19$ for
different redshifts are shown by lines (with redshifts as indicated in
the legend). The separation, $r$, is in comoving coordinates.  Filled
squares give the value of the correlation function of {\it dark matter} at $r=
5h^{-1}$Mpc at the same six redshifts ($z=0$ at the top, $z=5$ at the
bottom). \emph{Lower panel:} the ratio of the the galaxy correlation
function to the dark matter correlation function scaled by the
analytically derived linear bias term, for the same six redshifts (see
text for details).}
\label{fig:xievolve}
\end{figure}

\section{Images of the galaxy distribution}

Fig.~\ref{fig:sixslice} shows the distributions of dark matter and
galaxies in slices through the GIF N-body simulation volume at six
different redshifts. Each slice has dimensions $141.3 \times 141.3
\times 8 h^{-3}\hbox{Mpc}^3$. The dark matter is represented by the
greyscale, obtained by adaptively smoothing the N-body mass
distribution. The shade intensity is proportional to the logarithm of
the dark matter density in each pixel (the darker the pixel, the
greater the projected density of the slice). Each galaxy brighter than
$M_{\rm B}-5\log h = -19$ that formed in this region is depicted as a
circle whose size is proportional to the rest-frame B-band absolute
magnitude and whose colour indicates the rest-frame B-V colour, as
given by the scales at the top of the figure.  Fig.~\ref{fig:fourzoom}
displays zoomed images, at three redshifts, of the areas delineated by
the coloured boxes in Fig.~\ref{fig:sixslice}: the region around a
supercluster (left), and a more typical region (right).

\subsection{The biased galaxy distribution} 

The images in Fig.~\ref{fig:sixslice} illustrate what a realistic
distribution of galaxies might look like.  At $z=0$, this patch of
universe is populated by galaxies with a wide range of colours. The
dark matter has acquired the filamentary appearance characteristic of
gravitational growth from cold dark matter initial conditions.  The
galaxies ``light up'' the filaments and superclusters of the dark
matter distribution, but are conspicuously absent from regions in
which the density of dark matter is low. The emptiness of the voids is
quite striking (c.f. \pcite{peeblesvoids}). The voids would not be as
empty of galaxies as they are if, instead of using the semi-analytic
model, ``galaxies'' were simply identified with randomly selected dark
matter particles. Such a Poisson process inevitably places a small
fraction of galaxies in voids, but the physics of galaxy formation do
not allow this: at the magnitude limit appropriate to
Figs~\ref{fig:sixslice} and \ref{fig:fourzoom}, voids are truly voids:
no galaxies form in them.

While to the eye the distribution of galaxies in the figures may
appear to follow the broad features of the dark matter distribution,
it is impossible to judge in this way how faithful this tracing really
is. That galaxies in CDM models might not be perfect tracers of the
mass has been suspected for some time \cite{davis85,BBKS} and was
demonstrated explicitly for the GIF simulations by \scite{kauff99a}
and \scite{meetal}. These studies showed, for example, that on scales
below a few megaparsecs, galaxies in the simulations are less strongly
clustered than the dark matter (or ``anti-biased'' relative to the
mass). The upper panel in Fig.~\ref{fig:xievolve} shows the
correlation functions of the galaxies for the different redshifts
shown in Fig.~\ref{fig:sixslice} and demonstrates one aspect of the
bias. The correlation functions of these galaxies have remarkably
similar amplitudes, in contrast to the correlation function of dark
matter (whose value at $r=5h^{-1}$Mpc is shown as solid points in
Fig.~\ref{fig:xievolve}), the amplitude of which evolves rapidly with
redshift. In Fig.~\ref{fig:xievolve}, we also compare our
determinations of the correlation function to those derived from an
analytical approach to the bias (c.f. \pcite{baugh99}). For this, we
compute the effective bias of the galaxy population as:
\begin{equation}
b(z) = {1\over N} \sum_{i=1}^N b(M_i,z),
\end{equation}
where $N$ is the number of galaxies in the simulation, $M_i$ is the
mass of the halo in which the $i^{\rm th}$ galaxy is found, and
$b(M,z)$ is the bias of dark matter halos of mass $M$ and redshift
$z$, which we calculate using the fitting formula of
\scite{jing98} (which is based on the model of \pcite{mowhite96}). An
approximation to the galaxy correlation function is then
$b(z)^2\xi_{\rm DM}(r,z)$, where $\xi_{\rm DM}$ is the correlation
function of dark matter. (For the galaxy samples in our simulation, we
find $b(z)=1.07$, 1.24, 1.47, 2.05, 2.91 and 5.21 for $z=0.0$, 0.5,
1.0, 2.0, 3.0 and 5.0 respectively.)  In the lower panel of
Fig.~\ref{fig:xievolve} we plot the ratio of the galaxy correlation
functions in the simulation to the analytical approximation. On scales
above a few megaparsecs the analytical bias approach works well (it
does slightly underestimate the correlation functions, but given the
large biases present in our high-redshift samples, the approximation
is actually rather good), but, as expected, it fails on smaller scales
where our model predicts a scale dependent bias.

Certain kinds of bias are readily apparent to the eye in
Fig.~\ref{fig:sixslice}. For example, the largest dark matter clumps
at $z=0$ are preferentially populated by red galaxies, while the field
contains a mixture of galaxy colours. Similarly, the brightest
galaxies in the region are also preferentially found at the centres of
rich clusters. The first of these biases, the ``colour-density''
relation is intimately related to a ``morphology-density'' relation
and is a natural outcome of hierarchical clustering, as we will
discuss in more detail below. It has been investigated before in these
simulations, in a somewhat different form, by \scite{kauff99a} and
\scite{benson00c}, who found that the two-point correlation function
of elliptical galaxies is higher than that of spirals on small
scales. This kind of bias is, of course, known to occur in the real
universe (e.g. \pcite{davis76,loveday95}).

Biases in the distribution of galaxies are the inevitable by-product
of the complex physics of galaxy formation.  They affect
galaxies of different types in different ways and need to be
understood before attempting to interpret cosmological observations.

\subsection{The evolution of the galaxy population}

The evolution of galaxies is driven by a number of processes. The most
obvious one is the aging of the stellar populations. Even if galaxies
lived in isolation, stellar evolution would cause their luminosities
and colours to evolve.  However, no galaxy is an island: accretion,
mergers and interactions are common. Many galaxies are observed to be
forming stars today, possibly by converting new gas supplied
externally (although see
\pcite{bbfw00}), thus increasing their mass and size. Others are observed
to be expelling gas through galactic winds and tidal
encounters. Nucleosynthesis in stars causes the metallicity of
galactic gas and stars to evolve and this, in
turn, affects the integrated stellar spectra.

The nett effects of the various processes driving galaxy evolution are
readily apparent in Fig.~\ref{fig:sixslice}.  Since young stars
produce copious amounts of blue light, galaxies with high star
formation rates relative to their total stellar mass appear bluer than
galaxies with low relative star formation rates. At $z=0$, the
majority of the galaxies in the simulation are quite red because they
have low relative star formation rates and are made primarily of old
stars. As we look into the past, the appearance of the population
rapidly changes. By $z=1$, galaxies are much bluer because the typical
relative star formation rate is higher than at $z=0$ and the average
stellar age is younger. Beyond $z=1$, the galaxies remain blue,
reflecting the youth of their stellar populations. The apparent star
formation rate per unit volume declines at high redshift because fewer
galaxies are seen above our magnitude selection, even though those
that are seen still have high star formation rates. The detailed star
formation history in our model is discussed in \scite{coleetal}. It
is broadly in agreement with the star formation history of the
Universe as inferred from observations by, amongst others, 
\scite{lilly96}, \scite{madau96} and \scite{steidel98}.  Although many
quantitative details of the observations remain uncertain (due to
complications arising, for example, from dust obscuration), the
general behaviour seems to consist of a rapid rise in star formation
rate between $z=0$ and $z=1$, followed by a slowly declining (or
perhaps constant) star formation rate at higher redshifts. This is the
kind of behaviour exhibited by our simulations.

As the images in Fig.~\ref{fig:sixslice} illustrate, in hierarchical
models of galaxy formation the number of galaxies is constantly
changing. Galaxies are born as new dark matter halos form and gas is
able to cool in them and turn into stars. The population is depleted
when galaxies merge together. Of course, the number of galaxies
detected in a particular survey will depend crucially on the selection
criteria. All these effects can be clearly seen in the images of our
simulations. At $z=5$, there are very few galaxies present because
only a handful of massive dark halos have had time to collapse. In
those that have, galaxies have had little time to form stars, while
feedback from supernovae has strongly suppressed star formation in
small halos. The majority of the galaxies seen in the images at this
epoch occur in halos of mass $10^{11-12}h^{-1}M_\odot$ and have
stellar masses of a few times $10^9h^{-1}M_\odot$; the very brightest
galaxies are found in the tail of halos extending to masses close to
$10^{13}h^{-1}M_\odot$.  By $z=3$ the number of galaxies has increased
significantly, as more halos have collapsed and more galaxies have
been able to form. At this epoch several extremely bright (in the
B-band) galaxies are visible. The increased number of galaxies in the
image is due, in part, to our selection in the B-band which is
sensitive to the star formation rate. At $z=2$ there is a noticeable
increase in the abundance of galaxies as structures continue to
form. The most obvious change from $z=2$ to $z=1$ is a substantial
reddening of the galaxies, a trend which continues to $z=0$ as star
formation rates decline and stellar populations age.

Many faint blue galaxies are formed in the filamentary network of the
dark matter. In the comoving coordinates of Fig.~\ref{fig:sixslice},
galaxies move rather little between $z=5$ and the present. For
example, the progenitor of the large supercluster marked by a red box
at $z=0$ is already clearly visible at $z=3$ as a concentration of
young galaxies. In other words, by virtue of forming in the highest
density regions, galaxies are strongly biased at birth. This is a
fundamental outcome of hierarchical clustering
\cite{kaiser84,davis85}. It underlies the results of \scite{baugh98}
and \scite{governato98}, who argued that Lyman-break galaxies at $z=3$
would be expected to be strongly clustered, as was subsequently found
to be the case observationally \cite{adelberger98}.

\section{Quantitative analysis and comparisons with observations}

In this section, we compare the properties of our model with the limited
observational data currently available. More stringent comparisons will be
possible with the forthcoming 2dF and Sloan galaxy redshift surveys. We
consider, in turn, the variation of the clustering length with the
luminosity of a sample, the morphology-density relation and the topology of
the galaxy distribution.

\subsection{Correlation length versus separation}

\begin{figure}
\psfig{file=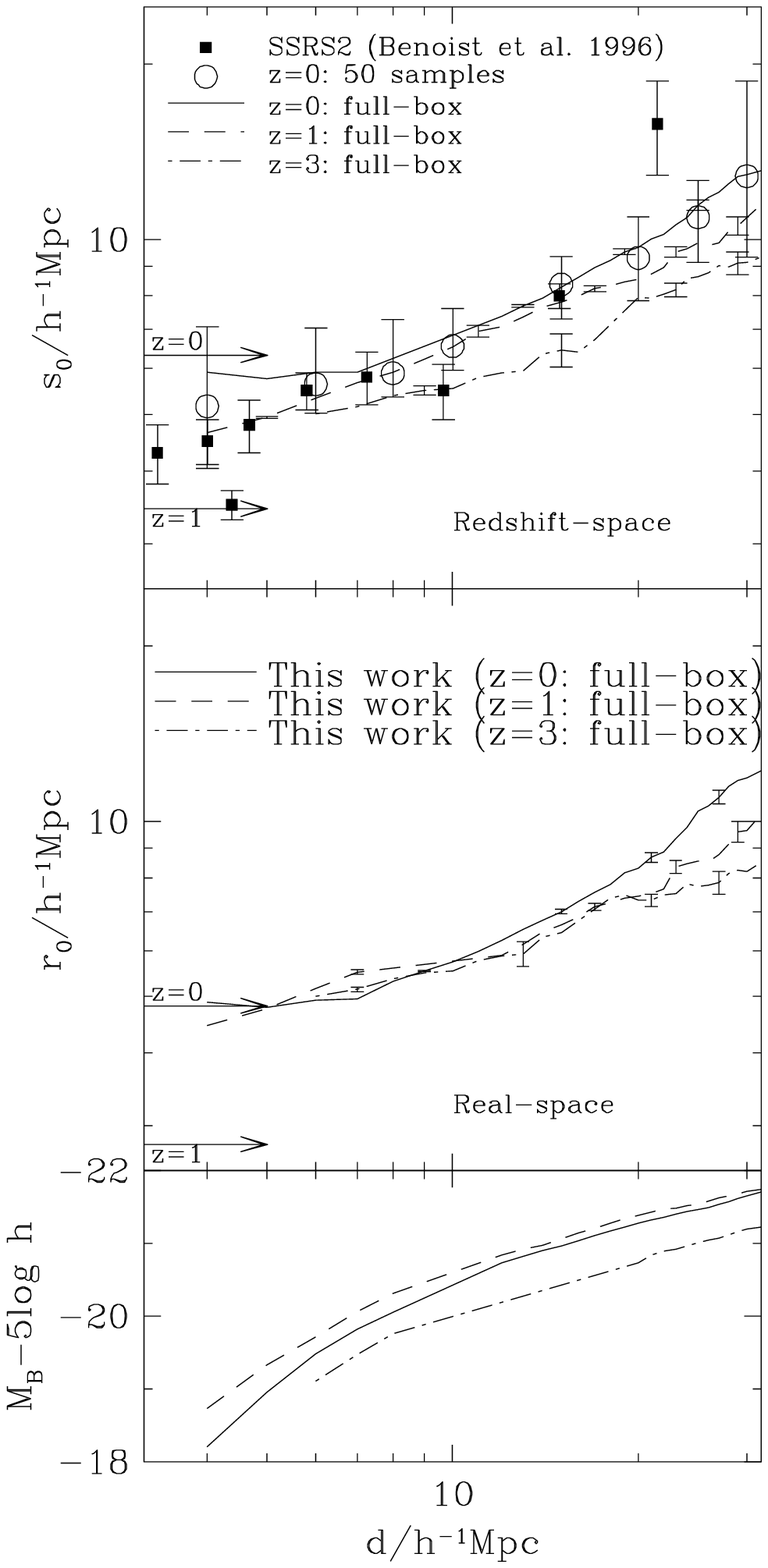,width=80mm,bbllx=5mm,bblly=55mm,bburx=110mm,bbury=265mm,clip=}
\caption{The redshift-space (top panel) and real-space (middle panel)
correlation lengths of galaxies as a function of the mean galaxy
separation (in redshift and real space respectively). Filled squares
in the top panel show the measured correlation lengths in the SSRS2
survey \protect\cite{benoist96}. The solid lines show our model
results for galaxies selected by their (dust-extinguished) B-band
magnitude at $z=0$, with error bars indicating the statistical
uncertainty. Open circles in the top panel show the median
redshift-space correlation lengths estimated from fifty subsamples
from the simulations, each of volume equal to that of a volume-limited
SSRS2 sample of the same absolute magnitude. Errorbars indicate the
10\% and 90\% intervals of the correlation length distribution. Dashed
and dot-dashed lines give our model predictions for galaxies at $z=1$
and $z=3$ respectively, selected according to their rest-frame B-band
magnitude. Horizontal arrows indicate the correlation lengths of the
dark matter at $z=0$ and $z=1$. The lower panel indicates the
dust-extinguished B-band absolute magnitude corresponding to a given
mean galaxy separation at $z=0$ (solid line), $z=1$ (dashed line) and
$z=3$ (dot-dashed line).}
\label{fig:s0vss}
\end{figure}

The evolution of the two-point correlation function of {\it dark
matter} in CDM models is now well established. In the linear regime,
it evolves according to the linear growth factor; in the non-linear
regime, its evolution can be calculated accurately using N-body
simulations (see, for example, \pcite{arj98}). By contrast, the
evolution of the two-point correlation function of {\it galaxies} has
only begun to be investigated in detail recently. In
Fig.~\ref{fig:s0vss} we plot the redshift-space correlation lengths,
$s_0$ (top panel), and the real-space correlation lengths, $r_0$
(middle panel), of galaxies brighter than a particular rest-frame
(dust-extinguished) B-band magnitude, as a function of their mean
separation, $d$, at $z=0$ (solid line), $z=1.0$ (dashed line) and
$z=3$ (dot-dashed line). For reference, the lower panel of
Fig.~\ref{fig:s0vss} shows the relation between mean separation and
absolute B-band magnitude. Statistical errors (obtained by assuming
Poisson counting statistics to estimate the error in $\xi(r)$ and
propagating this error through to the determination of $r_0$) are
shown by the error bars. At small pair separations, we show results
from the GIF simulation (applying the small correction for finite
volume effects discussed above), and at large pair separations, we use
the $512^3$ simulation which is more accurate on large scales. (The
sudden decrease in the size of the errorbars at $\approx 15h^{-1}$Mpc
is due to this change of simulation volume.) For reference, we show
the correlation lengths of the dark matter at $z=0$ and $z=1$ as
horizontal arrows (at $z=3$ the dark matter correlation length is
$1.1$ and $1.5h^{-1}$Mpc in real and redshift-space respectively).

\begin{figure*}
\centerline{\psfig{file=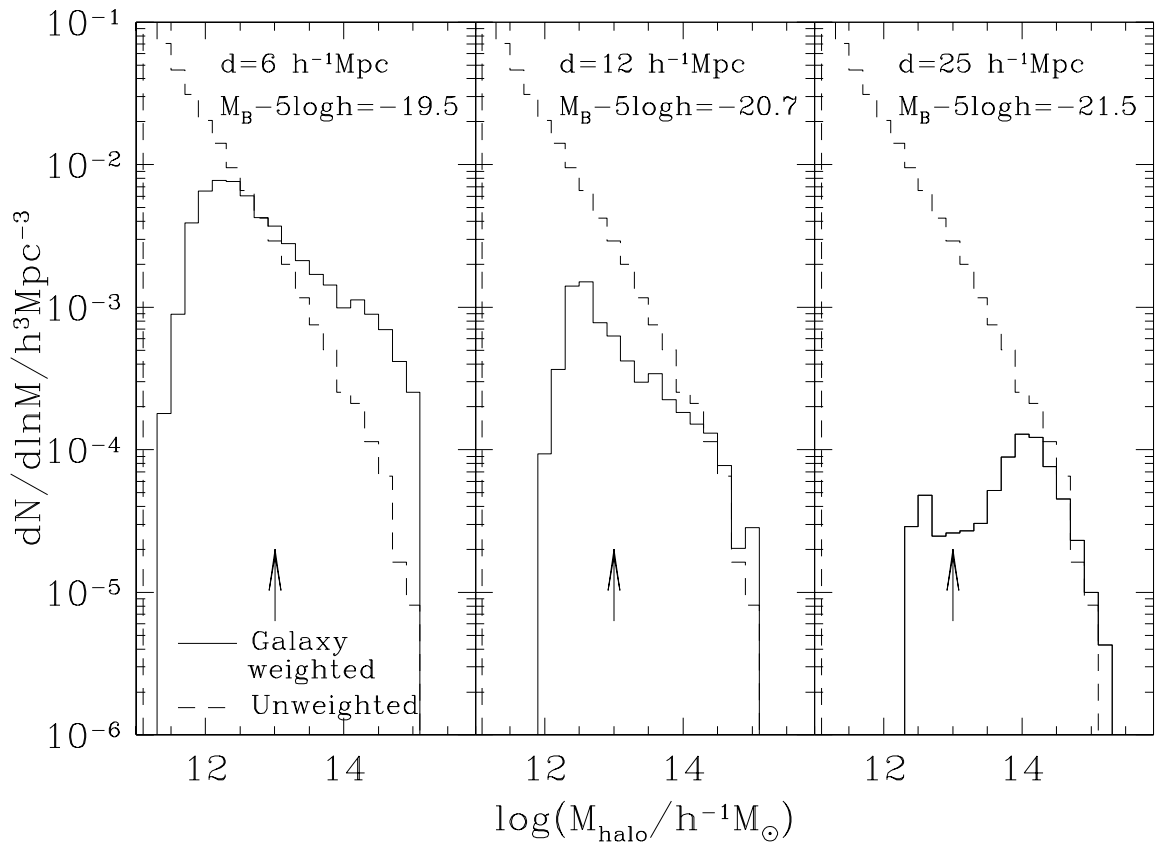,width=160mm,bbllx=0mm,bblly=125mm,bburx=190mm,bbury=265mm,clip=}}
\caption{Dark matter halo mass functions weighted by the number of
galaxies in each halo at $z=0$ (solid histograms). Results are shown
for galaxies brighter than three different dust-extinguished B-band
magnitudes, chosen to give a desired mean inter-galaxy separation,
$d$, as indicated in each panel. We show the (unweighted) mass
function of dark matter halos in each panel for comparison (dashed
histograms). The vertical arrow in each panel shows the location of
$M_*$ (defined by $\sigma(M_*)=\delta_{\rm c}$).}
\label{fig:ssmhalo}
\end{figure*}

We compare our model predictions with results from the Southern Sky
Redshift Survey 2 (SSRS2) \cite{benoist96}, which are shown by filled
squares in Fig.~\ref{fig:s0vss}. These data exhibit a nearly constant
redshift-space clustering length for galaxies fainter than $M_{\rm
B}-5\log h=-20$, followed by a rapid increase at brighter
magnitudes. \scite{benoist96} compared their data to two simple models
of galaxy bias. In the first, they assigned a dark matter halo mass to
galaxies in their sample using the Tully-Fisher and Faber-Jackson
relations, and then applied the techniques of \scite{mowhite96} to
compute the bias of these halos (and hence of the galaxies which
occupy them). This model provides a reasonable match to the observed
behaviour of the faint galaxies in the survey, but it is unable to
reproduce the strong luminosity dependent bias observed for galaxies
brighter than $L_*$. Their second model is based upon the work of
\scite{bernardeau92}, who developed a description of bias from the
non-linear evolution of the density field. This model is able to match
the luminosity dependent bias for bright galaxies, but predicts too
strong a relation for faint galaxies and so is also ruled out by the
data. Our model of galaxy clustering, on the other hand, does produce
a trend similar to that observed, namely a relatively constant
clustering length for small $d$, followed by a rise in correlation
length for the rarest objects. 
From Fig.~\ref{fig:s0vss} it appears that our model is not consistent
with the data 
over the whole range of separations. However, this discrepancy may
result from sampling variance in the observations, as we discuss below.
First, we explain how this trend arises in our models.

It is well known that dark matter halos are biased relative to the
underlying mass, with the most massive halos being the most strongly
clustered (e.g. \pcite{frenk88,mowhite96}). Thus, for the brightest
galaxies to be the most strongly clustered, it is necessary that they
should preferentially inhabit more massive dark matter halos than
those occupied by their lower luminosity counterparts. In
Fig.~\ref{fig:ssmhalo} we show mass functions of dark matter halos,
weighted by the number of occupant galaxies (solid lines) and
unweighted (dashed lines), for three values of $d$. In each panel, the
solid histogram gives the galaxy-weighted halo mass function when
galaxies are selected by their dust-extinguished B-band magnitude.The
vertical arrow in each panel shows the location of $M_*$, defined by
$\sigma(M_*)=\delta_{\rm c}$, where $\sigma(M)$ is the mass variance
in spheres containing a mass $M$ on average, and $\delta_{\rm c}$ is
the critical overdensity for collapse in the spherical top-hat
model. A simple understanding of the galaxy bias for each sample may
be gained from this figure. For the two smaller values of $d$, the
relative numbers of galaxies in highly-biased, cluster-sized halos
($M_{\rm halo}\gsim 10^{14}h^{-1}M_\odot$) and in weakly-biased,
galaxy-sized halos ($M_{\rm halo}\sim 10^{12}h^{-1}M_\odot$) are
comparable in the two samples. As a result, these two samples have
quite similar correlation lengths. However, for the sparser sample
with $d=25h^{-1}$Mpc, the relative number in cluster-sized halos is
much higher than in the two other cases. As a result, this sample has
a larger correlation length than the other two. (For this very bright
sample, we find that many galaxies occupying halos near the peak mass
of $M_{\rm halo}\sim10^{14}h^{-1}M_\odot$ have undergone a recent
burst of star formation.)

As noted above, our model is not in perfect agreement with the data of
\scite{benoist96}. The SSRS2 survey, however, covers a relatively
small volume and so sample variance may not be
negligible. \scite{benoist96} estimated the effects of sample variance
on their results and concluded that the luminosity
dependence of $s_0$ for faint (sub-$L_*$) galaxies could well be due to
sample variance, while that for brighter galaxies seemed to be a real
effect. We can estimate the size of this uncertainty directly from our
simulations, which cover a much larger volume than the real
survey. For a given value of $d$, we extract from the simulation fifty
randomly placed cubic regions of volume equal to that of a
volume-limited SSRS2 catalogue cut at the same absolute magnitude. We
then measure the correlation length in each of the fifty cubes. In
Fig.~\ref{fig:s0vss} we plot the median correlation lengths from the
cubes as open circles, with error-bars indicating the 10\% and 90\%
intervals of the distribution. (We note that even for the largest
value of $d$ shown, the number of independent SSRS2 volumes that fit
within the $512^3$ simulation cube is still reasonably large, $\sim 
20$, and so our estimates of sample variance at these separations
should still be accurate.) Evidently, sample variance in a survey the
size of the SSRS2 is large, and this can account for the differences
with our model.  At the smallest values of $d$, the median value of
$s_0$ from the subsamples is biased low relative to that measured in
the full simulation volume, because a large fraction of the clustering
signal comes from galaxies in and around a few large clusters and
these are often missing from small volumes cut out of the simulation
box. At larger separations, the main source of sample variance is the
low abundance of the brightest galaxies. Our estimate of sample
variance confirm the conclusion reached by \scite{benoist96}, namely
that the luminosity dependence of $s_0$ observed for sub-$L_*$
galaxies in the SSRS2 is due to sample variance, while that for
brighter galaxies is real.

The evolution of the redshift-space correlation lengths of galaxies of
different abundance is also illustrated in Fig.~\ref{fig:s0vss}. For
samples selected according to our specific criteria (i.e. according to
rest-frame B-band absolute magnitude), the correlation lengths at
$z=1$ and $z=3$ are slightly smaller than at $z=0$. This variation,
however, is much smaller than the variation in the correlation length
of the dark matter, as indicated by the horizontal arrows in
Fig.~\ref{fig:s0vss}. Thus, galaxies selected according to our
criteria are more strongly biased at $z=1$ and $z=3$ than at $z=0$,
even for quite small values of $d$. The images of the galaxy
distribution displayed in Fig.~\ref{fig:sixslice} show exactly how
this effect arises. The galaxies present at the highest redshift,
$z=3$, have formed in regions that are destined to become incorporated
into clusters or superclusters of galaxies by the present day. Such
regions are amongst the most overdense at high redshift, and so
small-scale density fluctuations tend to collapse earlier there than
in less overdense regions \cite{kaiser84,davis85,BBKS}. Forming as
they do in the most highly biased regions of the universe, galaxies at
high redshift naturally end up being strongly biased themselves. Our
model predicts a strong luminosity dependence of clustering length for
the brightest galaxies from $z=0$ to 3, an effect which is seen in
both real and redshift-space

\subsection{Morphology-density relation}

The existence of a correlation between galaxy morphology and local
environment is a remarkable feature of the galaxy~population.
\scite{dressler80} showed that the fraction of galaxies of different
morphological types is strongly correlated with the local galaxy
density: elliptical and S0 galaxies are found preferentially in high
density regions while spiral galaxies are found preferentially in low
density regions.

Early N-body simulations suggested that a morphology-density relation
is a natural outcome of hierarchical clustering from CDM initial
conditions \cite{fwed95dummy}. This was explicitly shown to be the case
by Monte-Carlo based semi-analytic modelling
\cite{kauff95,baugh96}. These calculations had no information on the
spatial distribution of galaxies and so the relation they established
is one between morphology and cluster mass, rather than between
morphology and galaxy density. The implementation of semi-analytic
techniques in high resolution N-body simulations allowed the radial
distributions of different kinds of galaxies inside large clusters to
be calculated for the first time \cite{springel01,oka01}. This work
has shown, for example, that galaxies of different colours are
spatially segregated within the cluster.
 
The emergence of a colour-density relation is clearly illustrated in
the images of Fig.~\ref{fig:sixslice}. Redder galaxies (with $B-V\gsim
0.7$) which, in our model, are primarily ellipticals or S0's (see
Fig.~12 of \pcite{coleetal}), are over-abundant in the most overdense
regions relative to the field (e.g. the large supercluster in the
middle of the left-hand edge, or the large cluster near the centre of
the bottom edge). The correlation between galaxy colour (or
morphology) and environment is a byproduct of the biases discussed in
the preceding subsection: the oldest, reddest galaxies form in the
highest density regions where the production of elliptical galaxies by
mergers is also favoured.

\begin{figure*}
\begin{tabular}{cc}
\psfig{file=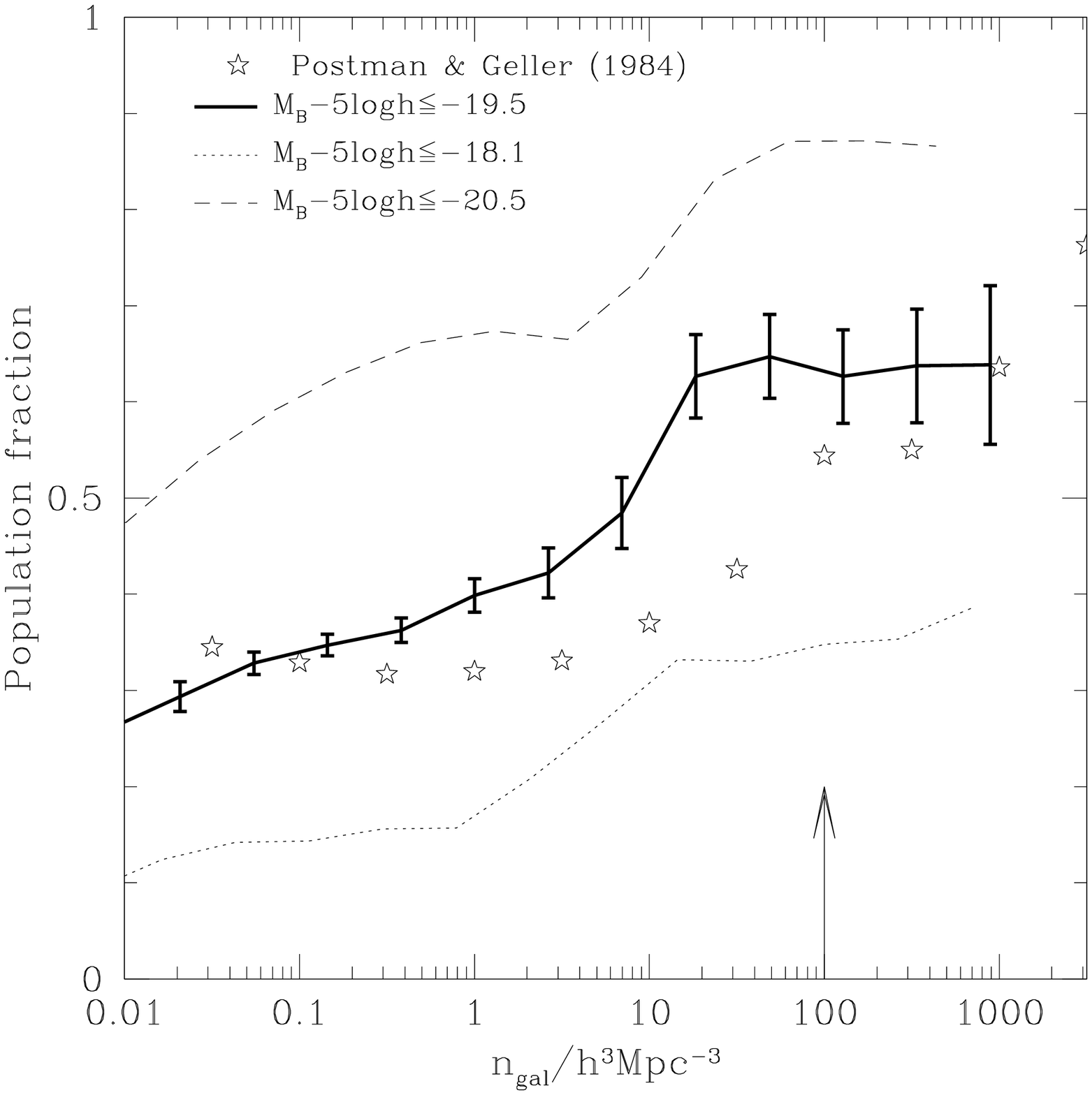,width=80mm} & \psfig{file=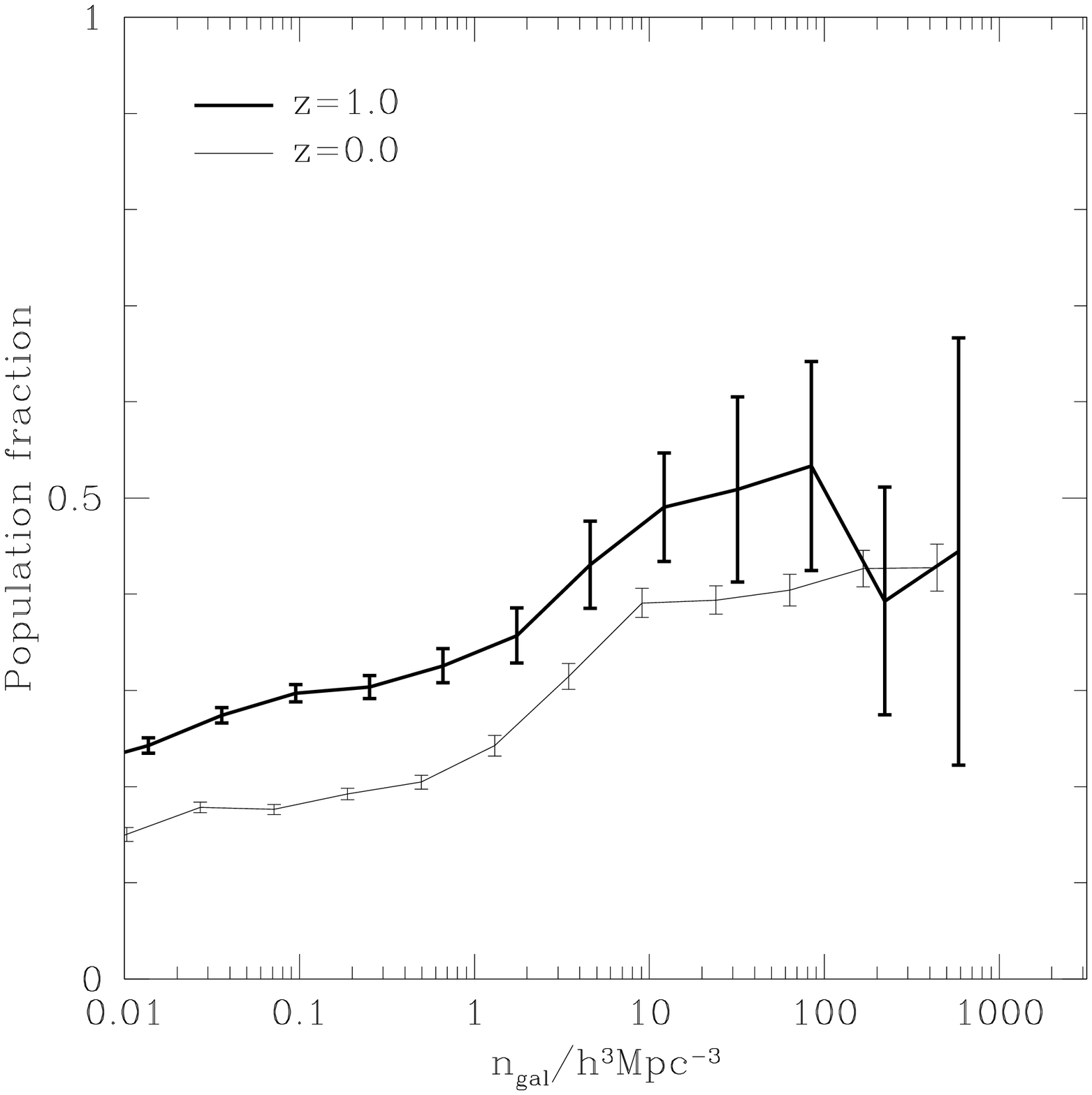,width=80mm}
\end{tabular}
\caption{The morphology-density relation for galaxies selected
according to their observed B-band magnitude. Stars show the fraction
of elliptical and S0 galaxies in the CfA redshift survey
\protect\cite{postman84}. We plot the group
fractions of \protect\scite{postman84} for densities less than
$100h^3$Mpc$^{-3}$ (as indicated by the vertical arrow), and their
cluster fractions for higher densities (where they define groups and
clusters as associations of 3-9 and $\geq$10 galaxies
respectively). In the left-hand panel, the heavy solid line is our
model prediction for the E/S0 fraction (i.e. galaxies with a
dust-extincted B-band bulge-to-total ratio, B/T$_{\rm B}>0.4$) at
$z=0$, obtained from the galaxy distribution in real-space. The
galaxies themselves are selected to have $M_{\rm B}-5\log h\leq-19.5$,
but the densities are extrapolated to $M_{\rm B}-5\log h=-17.5$ using
the CfA survey luminosity function of
\protect\scite{postman84}. Dotted and dashed lines are the
corresponding model results for galaxies brighter than the
completeness limits of the GIF and 512$^3$ simulations ($M_{\rm
B}-5\log h=-18.1$ and $M_{\rm B}-5\log h=-20.5$ respectively), and are
also corrected to $M_{\rm B}-5\log h=-17.5$. In the right-hand panel,
the heavy solid line shows the model result at $z=1.0$ for galaxies
above the simulation completeness limit, but here the densities are
left uncorrected and so correspond to $M_{\rm B}-5\log h=-18.4$. The
thin solid line shows the relation at $z=0$ for galaxies brighter than
$M_{\rm B}-5\log h=-18.4$ for comparison. Error bars are $1\sigma$
deviations.}
\label{fig:densmorph}
\end{figure*}

\begin{figure*}
\begin{tabular}{cc}
\psfig{file=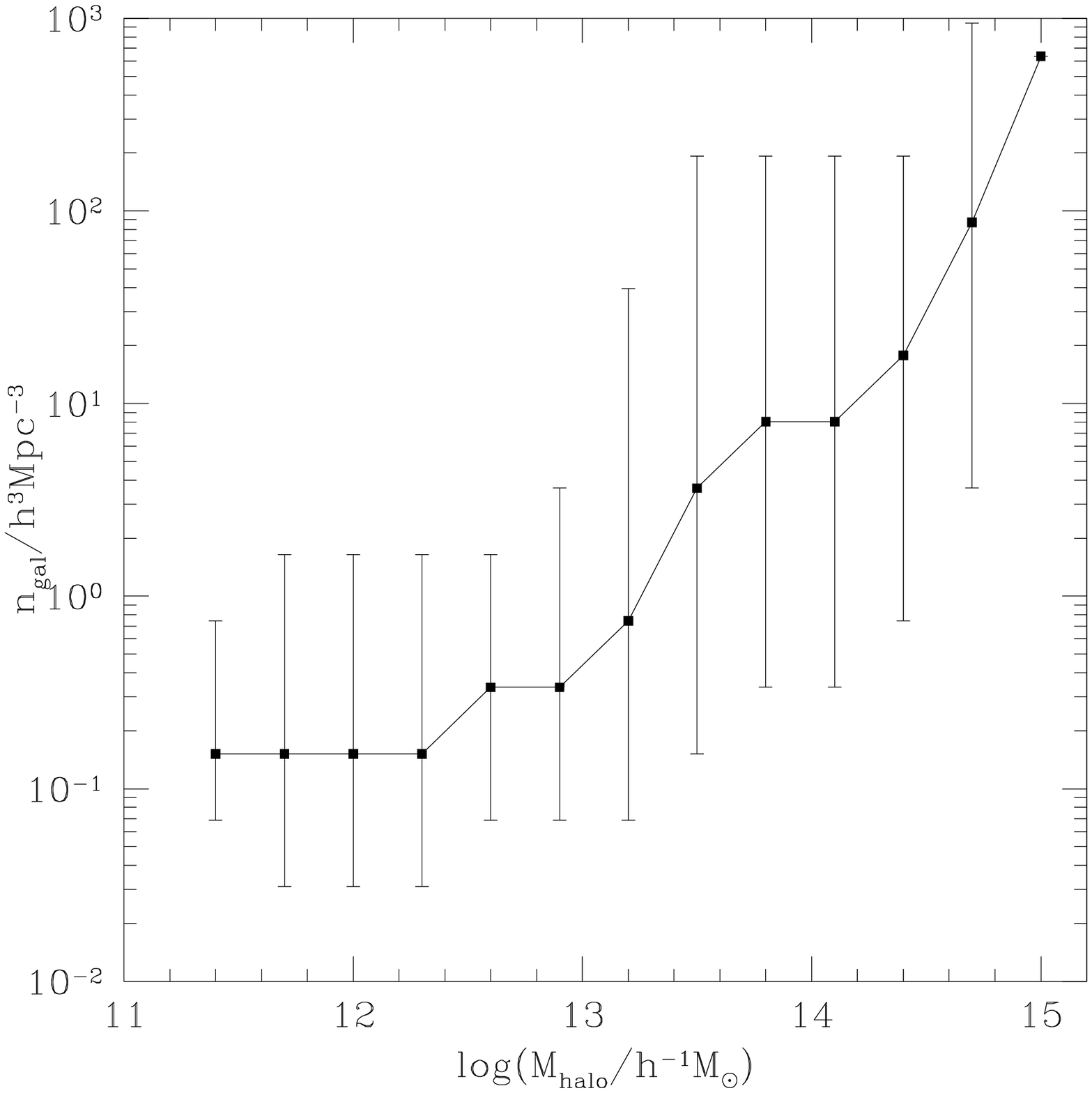,width=80mm} & \psfig{file=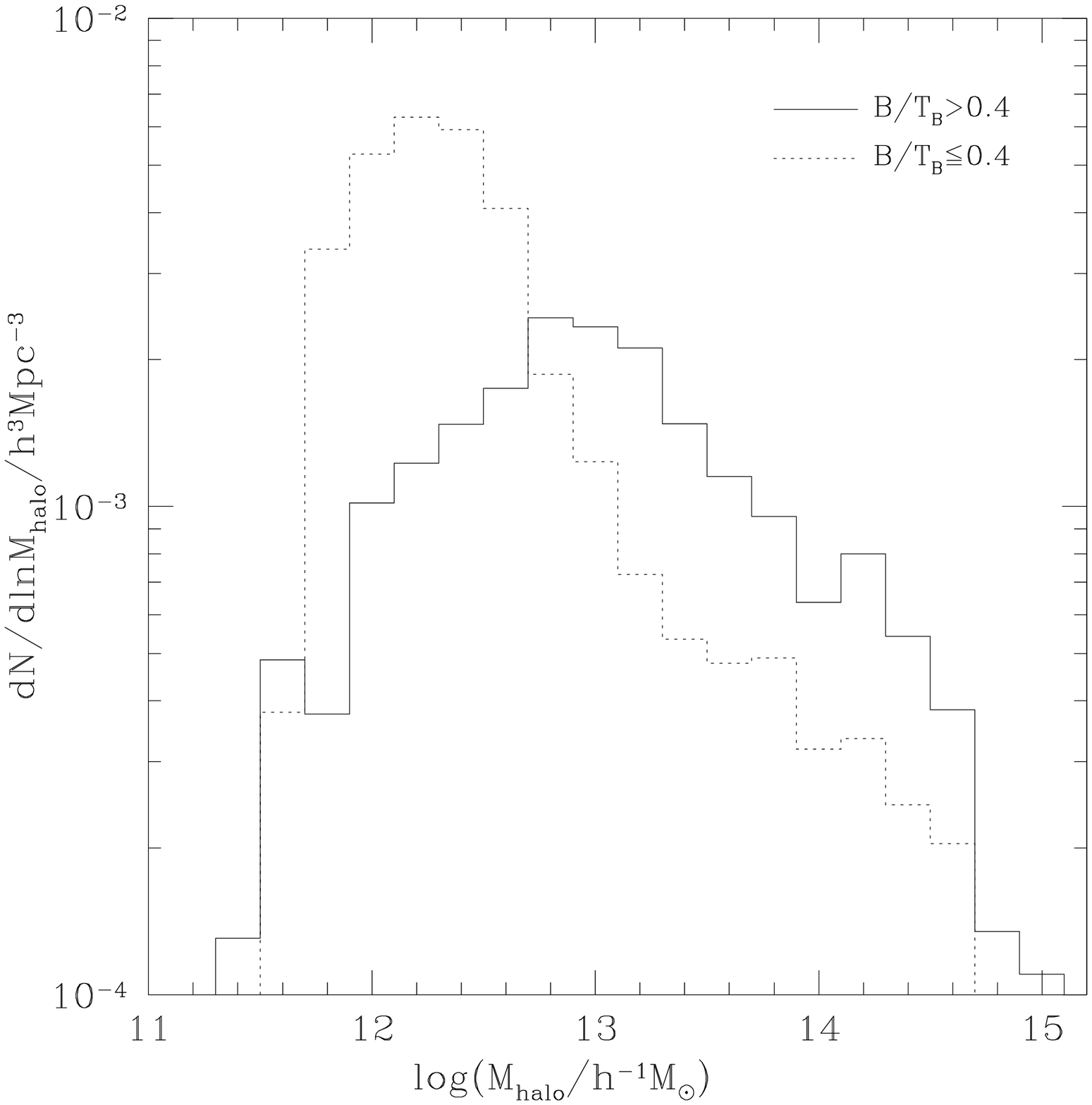,width=80mm}
\end{tabular}
\caption{\emph{Left-hand panel:} The local galaxy number density (for
galaxies brighter than $M_{\rm B}-5\log h=-19.5$, but corrected to
$M_{\rm B}-5\log h=-17.5$) as a function of dark matter halo
mass. Points show the median number density in each halo mass bin,
while error bars show the 10\% and 90\% intervals of the
distribution. \emph{Right-hand panel:} dark matter halo mass functions
weighted by the number of galaxies brighter than $M_{\rm B}-5\log
h=-19.5$ per halo. The solid histogram is the mass function for
galaxies with B/T$_{\rm B}>0.4$, while the dotted histogram shows the
mass function for galaxies with B/T$_{\rm B}<0.4$.}
\label{fig:mdhalos}
\end{figure*}

To quantify the morphology-density relation apparent in
Fig.~\ref{fig:sixslice}, we proceed in a manner analogous to the
analysis of an observational sample. We use a technique patterned on
that employed by \scite{postman84}. First, we apply a
friends-of-friends group-finding algorithm to the real-space
distribution of model galaxies brighter than a particular B-band
absolute magnitude, using different values of the linking
length. (\pcite{postman84} applied their group finder in
redshift-space, but used an anisotropic linking length to account for
distortions in the redshift direction. Since we have a galaxy
catalogue in real-space we perform group finding there, avoiding the
complications of redshift-space distortions.) In this manner, we build
up nested sets of groups as a function of the enclosed density (we
consider three or more galaxies linked together to be a
``group''). For very large linking lengths all galaxies will belong to
a group, but as the linking length is decreased each galaxy will at
some point no longer be a member of a group. We assign each galaxy a
local density corresponding to the surface density of the group of
which it was last a member. The local density at the surface of a
group formed with a linking length $r$ is approximately $n=3/2\pi r^3$
(e.g. \pcite{lc94}). The morphological type of the galaxy is assigned
according to our standard definitions based on dust-extinguished
B-band bulge-to-total luminosity ratios: galaxies with B/T$_{\rm
B}<0.4$ are labelled as spirals, while those with B/T$_{\rm B}\geq
0.4$ are labelled as E/S0 (\pcite{coleetal}). We constructed
density-morphology relations for samples with different limiting
absolute magnitudes. In each case, the measured densities were
corrected to the density of galaxies at $M_{\rm B}-5\log h=-17.5$
using the CfA survey luminosity function, as was done by
\scite{postman84}, namely we multiply the densities by a factor
\begin{equation}
f_{\rm n}={\int_{L_{\rm ref}}^\infty \Phi(L) \d L \over \int_{L_0}^\infty \Phi(L) \d L},
\end{equation}
where $L_0$ and $L_{\rm ref}$ are the luminosities corresponding to
$M_{\rm B}-5\log h\leq-19.5$ and $-17.5$ respectively and $\Phi(L)$ is
the luminosity function of the CfA survey.

The left-hand panel of Fig.~\ref{fig:densmorph} shows the model
morphology-density relation at $z=0$ for three absolute magnitude
cuts: $M_{\rm B}-5\log h=-18.1$ (dotted line; the completeness limit
of the GIF simulation), $-19.5$ (thin solid line; close to $L_*$) and
$-20.1$ (dashed line; the completeness limit of the $512^3$
simulation). It is immediately apparent that our model does display a
morphology-density relation with the correct trend: ellipticals/S0s
are more common in high density environments. For high densities
($\gsim 100$ galaxies $h^3$Mpc$^{-3}$) our model shows no relation
since, by construction, no morphology-density relation can exist
within individual halos because of the way in which we assign galaxies
to dark matter particles. (We do not plot model results for densities
greater than $1000 h^3$Mpc$^{-3}$ since these begin to probe single
dark matter halos in our simulation, resulting in a poor determination
of the morphology-density relation.) The right-hand panel of
Fig.~\ref{fig:densmorph} shows that our model predicts a very similar
morphology-density relation at $z=1$ as at $z=0$.

We can now elaborate a little on the cause of the morphology density
relation in our model, focusing on the morphology-density relation at
low densities (where the spatial distribution of morphological types
within individual halos is unimportant and the local density is
typically determined by averaging over regions containing many dark
matter halos). The morphological mix of galaxies in a single dark
matter halo can depend only upon the mass of that halo since this
statistically determines the merger history and galaxy formation
history in the halo. Therefore, for a morphology-density relation to
exist: (i) there must be a relation between dark matter halo mass and
local galaxy density and (ii) there must be a dependence of
morphological fraction on halo mass. The first of these requirements
is naturally met in hierarchical cosmologies since the most massive
halos are preferentially found in the densest environments. The
left-hand panel of Fig.~\ref{fig:mdhalos} shows the local galaxy
number density (for galaxies brighter than $M_{\rm B}-5\log h=-19.5$
but corrected to $-17.5$ as before) at the centres of dark matter
halos as a function of the halo mass (i.e. we plot the density of the
three-particle group, as defined above, which contains the central
galaxy). Clearly, the densest regions of the galaxy population are
associated with cluster-sized halos. The right-hand panel of
Fig.~\ref{fig:mdhalos} shows that our model also meets the second
requirement. This figure shows the dark matter halo mass function
weighted by the number of galaxies of a particular morphological class
per halo. The halo mass function for E/S0 galaxies is shifted to
higher mass halos relative to that for S galaxies. The resulting
dependence of morphological mix on halo mass merely reflects the fact
that galaxy formation in clusters is accelerated relative to the
field, allowing enough time for galaxy-galaxy mergers to produce a
large population of elliptical galaxies.

Fig.~\ref{fig:densmorph} also shows that the model morphology-density
relation depends upon the absolute magnitude at the which the galaxies
are selected. This property is simply a reflection of a
morphology-luminosity relation that is present in our model. This
property complicates the comparison with observational data since
existing analyses are usually based on apparent magnitude limited
samples.  In the figure we compare our model predictions with the
density-morphology relation measured in the CfA survey by
\scite{postman84}. For densities less than $100 h^3$Mpc$^{-3}$
(indicated by the vertical arrow), we plot the E/S0 fraction in
``groups'' (defined as associations of 3 to 9 galaxies by
\pcite{postman84}), and for larger densities, we plot the fraction in
clusters (associations of 10 or more galaxies). This shows that our
model displays qualitatively similar behaviour to the observational
data. A more detailed comparison may be possible with the new
generation of large redshift surveys.

Finally, another important prediction of our model is that there
should be a strong morphology-density relation well-established
already at $z=1$, as shown in the right-hand panel of
Fig.~\ref{fig:densmorph}. At this high redshift, the model relation is
qualitatively similar to that at $z=0$.

\subsection{The genus curve for the topology of the galaxy distribution}

The two-point correlation function contains only low order information
about the spatial distribution of galaxies. To fully specify this
distribution requires determining its higher order clustering
properties. Alternatively, the genus, a measure of the topology of a
smooth density field, provides a statistic that is sensitive to all of
the higher order moments of the distribution \cite{gott86,gott87}.

\begin{figure}
\psfig{file=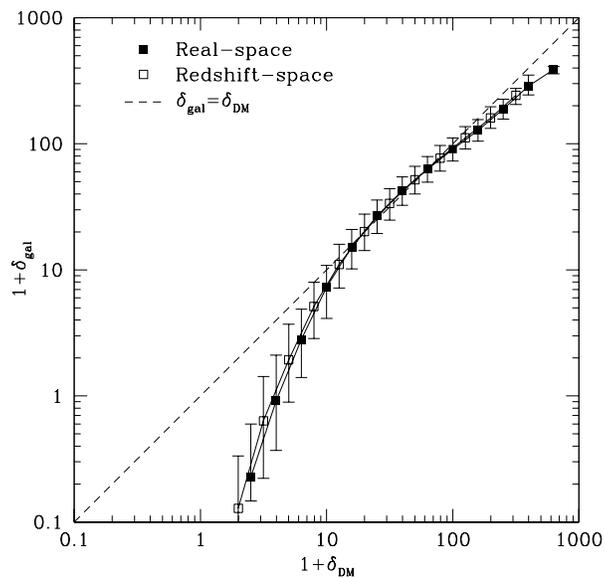,width=80mm}
\caption{The relation between the present-day galaxy and dark matter
overdensities in real-space (filled symbols) and redshift space (open
symbols). Only galaxies brighter than $M_{\rm B}-5\log h=-18.5$ are
considered and the two fields have been smoothed with the Gaussian
filter of eqn.~(\ref{eq:smoothfunc}), with $\lambda_{\rm e}=6.0
h^{-1}$Mpc. Points show the median galaxy overdensity at each dark
matter overdensity, and the error bars show the 10\% and 90\%
intervals of the distribution of $1+\delta_{\rm gal}$. When smoothed
on this scale redshift-space distortions make little difference to the
biasing relation.}
\label{fig:onepoint}
\end{figure}

The genus is defined as the number of topological holes minus the
number of isolated regions of an isodensity surface. By varying the
density at which this surface is placed, a genus curve can be
constructed. The genus curve has the interesting property that an
exact, analytic expression exists for the special case of a Gaussian
random density field, namely
\begin{equation}
g(\nu)=A(1-\nu ^2)\exp \left( - {\nu ^2 \over 2} \right),
\label{eq:genrp}
\end{equation}
where $g$ is the genus per unit volume and $\nu$ is defined by
\begin{equation}
\nu = \sqrt{2} {\rm erf}^{-1}(1-2f),
\end{equation}
where $f$ is the fraction of the volume above the density threshold
and ${\rm erf}^{-1}$ is the inverse of the error function. (For a
Gaussian random field, but not for any other field, this definition
implies that $\nu^2$ is the variance of the field.) The amplitude,
$A$, depends only on the second moment of the power spectrum of the
smoothed density field. In general, if the field is not Gaussian, the
shape of the genus curve may differ from eqn.~(\ref{eq:genrp}). With
the above definition for $\nu$, the genus curve remains the same under
any dynamical evolution or biasing in which the initial and final
densities at each Eulerian point are related by a monotonic,
one-to-one mapping.
Thus, this topological
measure provides, in principle, a method for testing whether or not
density fluctuations in the early universe were originally random and
Gaussian, as predicted in generic inflationary models.

The genus curve has been determined for both N-body simulations of
dark matter and for surveys of galaxies in the real Universe by many
authors (e.g.
\pcite{gott89,moore92,park92,rhoads94,vogeley94,canavezes97,springel98,canavezes00}.)
The genus curve for galaxies would be identical to that of the
underlying dark matter if the galaxy and dark matter density fields
were related by a monotonically varying factor, the bias. The bias
relation (i.e. the galaxy overdensity as a function of dark matter
overdensity) in our model is shown in Fig.~\ref{fig:onepoint} for
galaxies brighter than $M_{\rm B}-5\log h=-18.5$. It is qualitatively
similar to the relations found by \scite{somerville01a} using similar
techniques. The symbols show the median relation which is clearly
monotonic. However, there is substantial scatter around this relation
as a result of which it is no longer guaranteed that the genus curve
for galaxies will be identical to that of the dark matter.

\begin{figure*}
\psfig{file=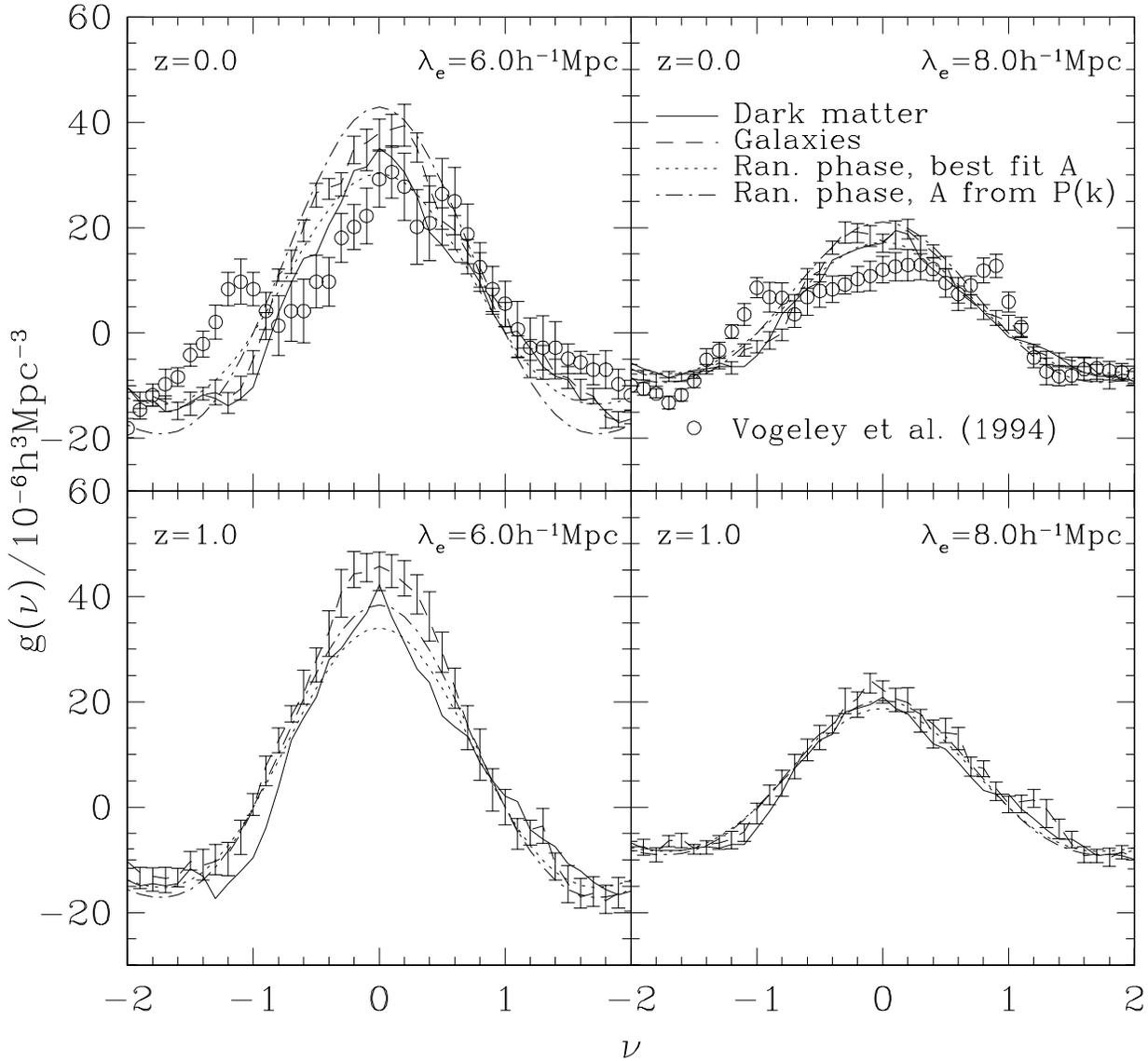,width=170mm,bbllx=0mm,bblly=90mm,bburx=190mm,bbury=265mm,clip=}
\caption{The genus per unit volume for two smoothing lengths,
$\lambda_{\rm e}=6.0$ and $8.0h^{-1}$Mpc, and two redshifts, $z=0$
and $z=1.0$, as indicated in the panels. Open circles show the results
from the combined CfA-I and CfA-II redshift surveys
\protect\cite{vogeley94}. Solid lines show the genus curves of dark
matter in our simulations, while the dotted lines show the random
phase genus curve which best fits the dark matter and the dot-dashed
line shows the genus curve for a Gaussian random field with the same
power spectrum as the dark matter. Dashed lines show the genus curves
of galaxies brighter than $M_{\rm B}-5\log h=-18.5$. All model curves
are calculated in redshift space. Errorbars on the model galaxy genus
curves are the standard deviations from thirty bootstrap resamplings
of the galaxy distribution. (Note, however, that the line indicates
the genus curve of the actual distribution, not the mean of the
bootstrapped samples.)}
\label{fig:genus}
\end{figure*}

We have used the technique described by \scite{coles96} to measure the
genus of both dark matter and galaxies in our simulation.
Fig.~\ref{fig:genus} shows the results for galaxies (dashed lines)
brighter than $M_{\rm B} - 5 \log h = -18.5$ at $z=0$ (upper panels)
and $z=1.0$ (lower panels) and also for dark matter (solid
lines). These curves were calculated by smoothing the dark matter and
galaxies in redshift-space onto a $128^3$ grid using a Gaussian filter
of the form
\begin{equation}
W(r)={1 \over \pi ^{3/2} \lambda_{\rm e}^3} \exp \left( - {r^2 \over
\lambda_{\rm e}^2} \right), 
\label{eq:smoothfunc}
\end{equation}
as is conventional in the literature on this subject. Smoothing
lengths of $\lambda_{\rm e}=6.0$ and $\lambda_{\rm e}=8.0h^{-1}$ Mpc
were chosen to match those used by \scite{vogeley94} in their analysis
of the CfA surveys. These smoothing
lengths are over five times larger than the size of the grid cells on
which the fields are tabulated and so the finite resolution of the
grid has no effect on the calculation of the genus curve
\cite{springel98}. Note that the mean comoving separation of galaxies
in our simulation is $4.3$ and $5.6 h^{-1}$ Mpc at $z=0$ and $z=1.0$
respectively (corresponding to 34000 and 16000 galaxies in the
simulation volume). At $z=1$ this is just smaller than the minimum
smoothing scale, thereby providing the greatest number of independent
resolution elements without allowing discreteness effects to become
too large \cite{weinberg87}. We estimate the errors on each genus
curve by bootstrap resampling of the galaxy catalogues. \scite{moore92}
find that this procedure produces slightly larger errors than those
estimated by considering several realizations of a mock catalogue.

In a $\Lambda$CDM cosmology, N-body simulations have shown that the
genus curve for the dark matter displays both a ``\emph{bubble
shift}'' (i.e. a shift to the right) with respect to the Gaussian
random phase genus curve and an amplitude reduction relative to a
Gaussian random density field having the same power spectrum
\cite{springel98}. These two effects can be seen by comparing the dark
matter genus curve to the two random phase genus curves (i.e. with the
shape given by eqn.~\ref{eq:genrp}) shown in Fig.~\ref{fig:genus} by
dotted and dot-dashed lines. The amplitude of the dotted curve is
chosen to best fit (in a least-squares sense) the measured dark matter
genus curve, while the amplitude of the dot-dashed curve is that
expected for a Gaussian random field with the same power spectrum as
the dark matter. The dark matter genus curve is displaced to the right
of the dotted curve, showing the bubble shift, and has smaller
amplitude than the dot-dashed curve, showing the amplitude drop. The
genus curve for galaxies brighter than $M_{\rm B} - 5 \log h = -18.5$
shows definite differences from that for the dark matter. Firstly, no
bubble shift exists for the galaxy genus curve which, instead,
exhibits a small ``{\em meatball shift}'' (i.e. shift to the
left). The galaxy curve also has a systematically larger amplitude
than the dark matter curve. \scite{canavezes97} advocate the amplitude
drop (defined as the ratio of the amplitudes of the best-fit random
phase genus curved for the actual density field and that of a Gaussian
random field with the same power spectrum) as a useful measure of the
degree of phase correlation in the galaxy density field. To measure
the amplitude drop, we ``Gaussianise'' the galaxy density field
(i.e. we take it to Fourier space, randomise the phases subject to the
reality condition, $\delta_k=\delta_{-k}^\star$, and then restore it
to real space). The amplitude drop is then simply the ratio of
amplitudes of the best-fit random phase genus curves for the original
and Gaussianized density fields. At $z=0$ we find amplitude drops of
$R=0.70$ and $R=0.84$ for the dark matter in redshift-space, for
$\lambda_{\rm e}=6.0$ and $8.0h^{-1}$Mpc respectively. (In real-space
we measure amplitude drops of 0.60 and 0.67 for the same two smoothing
lengths, in good agreement with the determinations of
\pcite{springel98}.) For galaxies at $z=0$ we find $R=0.84 \pm 0.02$
and $R=0.90 \pm 0.02$ also for these same two smoothing scales. At
$z=1$ (lower panels in Fig.~\ref{fig:genus}) the amplitude drops are
somewhat smaller, $R=0.90 \pm 0.03$ and $R=0.93 \pm 0.07$ for
$\lambda_{\rm e}=6.0$ and $8.0h^{-1}$Mpc respectively, as phase
correlations due to non-linear growth of structure have not had as
long to develop as at $z=0$.

As noted above, galaxies would have exactly the same genus curve as
the dark matter if there were a one-to-one mapping between dark matter
and galaxy density fields which preserved the density
ranking. However, we do see significant differences between the galaxy
and dark matter genus curves. This must be due either to the scatter
in the biasing relation between galaxies and dark matter, or to
systematic biases arising from the relatively small number of galaxies
$(\sim 10^4)$ compared to dark matter particles $(\sim 10^7)$ in our
samples. To test this latter possibility, we extracted twenty random
samples of dark matter particles with the same abundance as the
galaxies in our catalogue and computed their genus curves. We find
that this sparse sampling is the primary cause of the differences
between the galaxy and dark matter genus curves. Just as for the
galaxy sample, the sparsely sampled dark matter shows no evidence for
a bubble shift (and agrees closely with the galaxy genus curve for
$\nu <0$) and also shows a higher genus curve amplitude compared to
the fully sampled dark matter distribution. We conclude therefore that
any differences between the genus curves for dark matter and galaxies
due to the stochastic bias of Fig.~\ref{fig:onepoint} are negligible
compared to the effects of sparse sampling of the galaxy density
field. This sparse sampling at present severely limits the usefulness
of the genus statistic for quantifying the Gaussianity of the initial
dark matter distribution.

We also show in Fig.~\ref{fig:genus} the genus curve measured for the CfA
surveys by \scite{vogeley94}. While our model results are in
reasonable agreement with these data, the pronounced features in the
data suggest that the CfA surveys do not have a sufficiently large
volume to avoid significant sample variance effects. Indeed, when we
extract ``CfA survey'' volumes from random locations in our
simulations and measure their genus curves, we find that excursions
such as those seen in Fig.~\ref{fig:genus} are very common.

\section{Discussion and conclusions}

We have implemented a semi-analytic model of galaxy formation in high
resolution N-body simulations of the $\Lambda$CDM cosmology in order
to study the spatial distribution of galaxies and its evolution. The
semi-analytic model requires setting values for a number of parameters
which describe the physical processes that are modelled, such as gas
cooling, star formation and the associated feedback mechanisms and
chemical evolution, galaxy merging, the evolution of stellar
populations, etc. In keeping with the general philosophy of our work
in the subject, we have fixed all the model parameters by requiring a
match to a handful of global properties of the local galaxy
population, with the largest number of constraints coming from the
local B-band and K-band luminosity functions. No further adjustment to
these parameter values was allowed in the clustering study carried out
in this paper. Thus, our clustering results are genuine predictions of
the model and offer an opportunity to test the validity of the
physical assumptions it requires as well as the realism of the
$\Lambda$CDM model as a whole. In this paper, we have considered three
specific statistical measures of clustering: the correlation length
(in real- and redshift-space) of samples of galaxies of different
luminosity, the morphology-density relation and the genus curve.  At
$z=0$, our model may be tested by forthcoming data from the 2dF and
Sloan surveys; at $z=1$, it may tested by planned surveys such as DEEP
\cite{davis98} and VIRMOS \cite{lefev99}.

The results presented here extend and complement those presented in
earlier papers in this series \cite{meetal,benson00c}, as well as in
the series of papers by Kauffmann and collaborators
\cite{kauff99a,kauff99b,diaferio99} who analysed one of the N-body
simulations that we have analysed here, but using their own
semi-analytic model. In \scite{meetal}, we examined the physical and
statistical processes that segregate the galaxies from the dark matter
and we showed that the (real-space) two-point galaxy correlation
function in a $\Lambda$CDM model that produces an acceptable galaxy
luminosity function is in remarkably good agreement with
observations. In \scite{benson00c}, we considered clustering in
redshift space and analysed the resulting distortions of the two-point
correlation function, as well as its dependence on galaxy luminosity,
morphology and colour. Kauffmann and collaborators studied many of
these properties too, at the present day \cite{kauff99a}, and at high
redshift \cite{kauff99b}, as well as the clustering of groups
\cite{diaferio99}. On the whole, these two independent analyses agree
quite well and the differences that do exist can be readily understood
in terms of differences in the detailed assumptions for the physics of
galaxy formation (see \pcite{benson00c} for a detailed discussion of
differences between the two models). The evolution of clustering has
also been studied using similar semi-analytic/N-body techniques by
\scite{governato98} and, more recently, by \scite{wechsler01}.

Images of our simulation clearly illustrate many of the salient
features of galaxy growth by hierarchical clustering. They show that
galaxies approximately trace the filamentary structure and avoid the
lowest density regions of the dark matter distribution, that the
redder galaxies tend to predominate in the most massive dark halos and
that the brightest galaxies occur almost exclusively in regions of high
dark matter density. A series of time slices shows how the galaxy
population changes in abundance and colour with the passage of time
and demonstrates the primary effect behind biased galaxy formation:
the formation of the first bright galaxies in regions of exceptionally
high dark matter density.

From quantitative studies of the galaxy distribution we reach the
following three conclusions:
\begin{enumerate}
\item The correlation length of galaxies in real and redshift-space
increases rapidly with galaxy luminosity for galaxies brighter than
$L_*$, both at $z=0$ and at high redshifts, $z\lsim 3$.
\item A strong morphology-density relation, in the
same sense as observed, is a natural outcome of hierarchical clustering
from CDM initial conditions, and radpidly develops in our simulation. A
clear morphology-density relation is predicted to be already in place at
least since $z=1$.
\item The topology of the galaxy distribution, as
measured by the genus statistic, differs significantly from that of
the dark matter. However, the differences are due almost entirely to
sparse sampling effects; the stochastic biasing between galaxies and
dark matter is, at most, a minor effect.
\end{enumerate}
We now discuss these points more detail.

The variation of the correlation length with luminosity or,
equivalently, with mean inter-galaxy separation is one of the most
striking results of our analysis. The correlation length is virtually
insensitive to the mean separation within the sample out to
separations of around $10h^{-1}$Mpc (corresponding to luminosities of
$M_{\rm B}-5\log h\approx-20.5$), but for brighter, sparser samples it
increases very rapidly. Thus, the redshift-space clustering length of
galaxies of luminosity $7 L_*$ is predicted to be over twice as large
as that of $L_*$ galaxies. The pattern is similar in real- and
redshift-space.  Owing to the large volume of our simulations, this is
the first time that this rapid increase in correlation length at the
brightest luminosities has been unambiguously demonstrated.  The main
cause of this behaviour is the preponderance of such galaxies at the
centres of massive clusters. These galaxies experience enhanced merger
rates at early times and are the beneficiaries of late accretion of
cool gas which, in our model, is always funnelled onto the central
galaxy in each halo. Our predictions for the dependence of clustering
strength on galaxy luminosity are in broad agreement with existing
datasets, but these are rather small and thus subject to considerable
sampling uncertainties. A better test of these predictions should be
forthcoming shortly from the 2dF and Sloan surveys.

Our simulations develop a strong morphology-density relation similar
to that observed in the local universe: elliptical/S0 galaxies
predominate in rich clusters while spirals predominate in the
field. We have used a technique patterned after observational
procedures to characterise the morphology-density relation in our
simulations and find that it quantitatively agrees rather well with
observations.  The cause of the morphology-density relation is closely
related to the reasons behind the luminosity dependence of
clustering. In rich clusters, galaxy evolution is ``accelerated''
relative to more ordinary regions of space, thus allowing sufficient
time for mergers and interactions to build up a large population of
bright elliptical/S0 galaxies. Remarkably, a strong morphology-density
relation is already well-established by $z=1$.

Finally, we have investigated the topology of the galaxy distribution,
providing the first theoretical prediction for the genus curve of {\it
galaxies}, rather than merely of {\it dark matter}. Of course, if galaxy
density were monotonically related to dark matter density, the two
curves would be the same. In the simulations there is on average a monotonic
relation between the two, but it has such large scatter that it does
not preclude differences in the respective genus curves. We do
actually find a difference in our simulations: the dark matter genus
curve has a ``bubble'' shift whereas the galaxy genus curve has a
``meatball'' shift and also a higher amplitude. It turns out, however,
that the differences are not due to genuine topological differences
but rather to the sparse sampling of the density field provided by
galaxies. The confusing effects of sampling have been pointed out by
previous authors (e.g \pcite{canavezes97,springel98}). Our simulations
show that, in order to measure an unbiased genus curve for a clustered
galaxy distribution (particularly at low overdensity), several hundred
galaxies per smoothing volume are required. This is unlikely to be
practical even with the new generation of large surveys and so we will
have to live with these biases. The best approach for comparing models
and data is therefore to analyse each in identical ways, so as to
cancel out any systematic effects. Mock galaxy catalogues such as
those presented here will be crucial for this approach.

To summarise, the combination of high-resolution N-body simulations of
dark matter and semi-analytic modelling of galaxies provides a
powerful technique for turning cosmological and galaxy formation
theory into realistic realizations of the galaxy population that can
be compared in detail with observations. Tests of this sort will
become increasingly common with the new generation of galaxy
surveys. In this way, it will be possible to extract not only the
cosmological information encoded in the clustering pattern, but also
valuable information regarding the physics of galaxy formation.

\section*{Acknowledgements}

AJB, SMC and CSF acknowledge receipt of a PPARC Studentship, Advanced
Fellowship and Senior Fellowship respectively. CSF also acknowledges a
Leverhulme Research Fellowship.  CGL acknowledges support at SISSA
from COFIN funds from MURST and funds from ASI. This work was
supported in part by a PPARC rolling grant, by a computer equipment
grant from Durham University and by the European Community's TMR
Network for Galaxy Formation and Evolution. We acknowledge the Virgo
Consortium and GIF for making available the simulations used in this
study.

\end{document}